%\magnification=1200 
\vsize=8truein 
\hsize=6truein 
\hoffset=18truept
\voffset=24truept

\font\smc=cmcsc10
\font\smallsmc=cmcsc8
\font\smallrm=cmcsc8

\headline={\ifodd\pageno \ifnum\pageno>1 \smallrm \hfil 
ABELIAN SOLITONS               % running head for right-hand page is title in 
caps
\hfil\folio \else\hfill\fi \else \smallrm \folio \hfill
RON Y. DONAGI AND EMMA PREVIATO   % running head for left-hand page is authors 
in caps
\hfill\fi} \footline={\hss}   % footline is blank

\vglue .5in

\centerline{\bf
ABELIAN SOLITONS               %use capital letters for title
}
\vglue .5in
\centerline{\smc        %cap and lowercase for author
Ron Y. Donagi and Emma Previato
}
\vglue .5in
\narrower{\noindent
{\smc Abstract.}
We describe a new algebraically completely integrable system,
whose integral manifolds are co-elliptic
subvarieties of Jacobian varieties.  This is a multi-periodic
extension of the Krichever-Treibich-Verdier system, which
consists of elliptic solitons.
}
\vglue .5in
\footnote{}{\hskip-\parindent

Ron Donagi is partially supported by NSF grant DMS-9802456.
Emma Previato gratefully acknowledges partial
support under NSF grant DMS-9971966.}

The goal of this work is to generalize the theory of elliptic solitons,
which was developed by A. Treibich and J.-L. Verdier based on earlier works
by H. Airault, H.P.
McKean and J. Moser [AMcKM], and I.M. Krichever [K]. Elliptic solitons are a
subclass of algebro-geometric KP solutions (cf. {\bf \S 1}),
namely those that are elliptic functions in the $x$ variable.
Viewed as
Jacobians which contain special configurations of elliptic curves, the
elliptic solitons are dense in the classical topology [CPP].

We pose the problem of describing the moduli of abelian solitons, that is,
KP-solitons whose first $k\geq 1$ 
%(nontrivial) 
flow variables are tangent 
to some proper abelian subvariety of the Jacobian.  We explore several
possible constructions of such abelian solitons. Most of these actually
fail, cf. the appendix. We do succeed in constructing, for any natural
number $k$ and elliptic curve $C$, an infinite sequence of families of
co-elliptic solitons, which are periodic along $k$-dimensional abelian
subvarieties of ($k+1$)-dimensional Jacobians, with quotients isogenous to
the given elliptic curve $C$.  We show (Theorem 2.1)
how these families of co-elliptic
solitons organize into algebraically integrable systems (aci), which can
be identified as nonlinear subsystems of Markman's system of meromorphic
Higgs bundles.       

\bigskip
\noindent {\bf \S 1 Abelian solitons}
\medskip
We fix a smooth point $p$ on a curve $\widetilde{C}$
of genus $g$, use another smooth
point $p_0$ to define the Abel map $A_{\widetilde{C}}$ 
on the smooth part of $\widetilde{C}$ (but this choice is 
not intrinsic to the problem)
and study the image of a neighborhood of $A_{\widetilde{C}}(p)$ in 
Jac$\widetilde{C}$.
This allows us to define the
$g$ vector fields which are the flows
of the KP hierarchy, cf. e.g. [SW] for a definition of the hierarchy. 
\medskip
{\bf 1.1 The osculating spaces.} We
describe these both with and without choice of a local parameter near
$p$. Such a parameter is traditionally denoted by $z^{-1}$, with
$z^{-1}(p)=0$, and extends through the disk $|z^{-1}|\leq 1$ [SW]. 
%$\underline{\rm With\ choice.}$
Let $\omega_1,\ldots,\omega_g$ be a basis of
$H^0(\widetilde{C},\Omega_{\widetilde{C}} )$ and {\hfil\break}
$\omega_i =\sum_{j=0}^\infty
a_j^{(i)}z^{-j}d(z^{-1})$ the local expansion near $p$. The vectors
$U_{j+1}=(a_j^{(1)},\ldots,a_j^{(g)})$ osculate the curve ${
A}(\widetilde{C} )$ at $A(p)$ to order $j+1$, for instance $U_1$ is a
tangent vector since $A(q)=(\int_{p_0}^q\omega_i)_{1\leq i\leq g}$.
They generate translation-invariant vector fields on Jac$(\widetilde{C})$
which are called the ``KP flows.'' 
Without choosing $z^{-1}$, the curve
and the point only determine the flag of hyperosculating
vector spaces $<U_1>\subset
<U_1,U_2>...\subset{\bf T}:=
H^0(\widetilde{C},\Omega_{\widetilde{C}}^1)^\ast$.

Our goal will be to `integrate' the first $k$ flows in the sense of
finding a $k$-dimensional abelian subvariety whose tangent space
at every point contains $<U_1,U_2,...,U_k>$.

{\bf 1.2.0 Elliptic solitons.} By definition [TV1],
these are KP solutions which are elliptic in the first variable;
this means that the 1-parameter subgroup
generated by $U_1$ is an elliptic curve
$C$.

{\bf 1.2.1 Tangential covers.}
Let $\widetilde{C}, C$ be (projective, integral) curves of (arithmetic) genus
$>0,\ p\in\widetilde{C}$, $q\in C$ smooth points and $\pi :(\widetilde{C},
p)\rightarrow (C,q)$ a finite pointed morphism.
The choice of a smooth point in $\widetilde{C}$ and $C$ determines Abel
maps $A_{\widetilde{C}}$ and $A_C$; if $\pi^*$ denotes pull-back of line 
bundles,
the composition $\pi^*\circ A_C$ sends $C$
to Jac$(\widetilde{C})$. 
%Since $A_C$ is an isomorphism, it is customary
%to identify $C$ and $A_C(C)$.  
Then,  $\pi$ is called a
tangential cover if $\pi^*\circ A_C (C)$ and $A_{\widetilde{C}} (\widetilde{C}
)$ are tangent at the origin of Jac$(\widetilde{C})$.

[TV2] shows that elliptic solitons correspond  to tangential
covers of an elliptic curve (though not in a one-to-one
manner; extra data are needed in both directions). 
\medskip

{\bf 1.3.0 Abelian solitons.} Motivated by 1.2, we say that $\widetilde{C}$ 
gives
rise to a $k$-abelian soliton if Jac$(\widetilde{C})$ contains an abelian
subvariety $P$ of dimension $k<g$ and the flows $U_1,\ldots,U_k$
are tangent to $P$.
\medskip
 {\bf 1.3.1 Lemma.} {\it A covering $\pi\colon (\widetilde{C},
p)\rightarrow (C,q)$ with a ramification point of index $g-h+1$
at $p$, where $h=$ genus $C$, 
gives rise to a $(g-h)$-abelian soliton for $P=${\rm Prym}($\pi$).
More
generally, {\rm Jac}$(\widetilde{C})\sim A_1\oplus A_2$ with $A_i$ an abelian
subvariety of dimension $k_i$ is an abelian soliton with respect to
$A_2\Leftrightarrow \pi_{A_1}$ of $A_{\widetilde{C}}$ has a ramification point 
of
index $k_2+1$ at the origin.}
\medskip
Note that we adopted [H]'s terminology, according
to which the ramification index is the local-ring valuation
$e_p(\tau )$ where $\tau$ is a local parameter at $q$, so that
there is ramification exactly when the index is $>1$.
This lemma is simply a restatement of condition 1.3.0 on the
osculating flag (1.1) at $A_{\widetilde{C}} (p)$ and $A_C(q)$ under the 
pull-back
$\pi^\ast$. Indeed,
Jac$(\widetilde{C})\sim$ Jac$(C)\oplus P$, where $P=$Prym($\pi$) is the
Prym variety of the cover, namely the connected component
through the origin of the kernel
of the norm map. Again up to isogeny,
we call $\pi_1$ and $\pi_2$ the projections to Jac$(C)$ and $P$, respectively.
Then the osculating flag to $A_{\widetilde{C}} (\widetilde{C} )$
 is tangential to $P$ $\Leftrightarrow$ 
the differential of the projection $\pi_1$  vanishes on the
flag 
%tangent vector to $A_{\widetilde{C}} (\widetilde{C} )$ 
$\Leftrightarrow$ the
projection $\pi_1$ restricted to $A_{\widetilde{C}} (\widetilde{C} )$ is 
ramified at
the origin; and an abelian soliton for $P$ $\Leftrightarrow$
$\pi_1$ has a ramification point of index dim$P+1$ at the origin.
This situation is somewhat `dual' to the elliptic soliton 
$\widetilde{C}\rightarrow
C$, which indeed is unramified at $p$.

\medskip

{\bf 1.4 Coelliptic solitons.}  
Abelian solitons constructed from covers
seem to be quite rare, due to numerical constraints,
cf. {\bf Appendix}.
Here are the families we
could find. They give connected components of
the ``moduli space'' of abelian solitons. 

\medskip
{\bf Construction}. Fix integers $g \geq 2$ and $n \geq g$,
and an elliptic curve $C$. The Hurwitz formula implies that a map
$\widetilde{C} \to C$ of degree $n$
will be ramified at a total of $2g-2$ points (counting multiplicities),
where $g$ is the genus of $\widetilde{C}$. For the coelliptic solitons,
we bring $g-1$ of these ramification points together, i.e. we consider
the family of all covers $\widetilde{C} \to C$ of degree $n$ with one
ramification point where $g$ sheets come together, and generically $g-1$
additional simple ramification points. Note that by Riemann's 
existence theorem (cf. e.g. [Mu p. 15])
such a cover is determined, up to
finite choices, by specifying the location of the $g$ branch points
(=images of the ramification points) in $C$. Since the elliptic curve
$C$ has a $1$-parameter group of automorphisms, we can fix the location
of the multiplicity $g-1$ branch point at the origin $q \in C$. Our
remaining family of covers is $g-1$ dimensional: it is quasi-finite over the
$(g-1)$-st symmetric product of $C$. By Riemann's existence
theorem, any such cover is algebraic. In fact, it is an abelian soliton: by
Lemma 1.3.1, $\widetilde{C}$ osculates $P$ to order $g-1=$
dim $P$ where $P$ is the Prym variety of the cover.

\medskip
{\bf Remark.}  
 The same idea won't give higher-than-coelliptic
abelian solitons, for if $\widetilde{C}\rightarrow C$ is an $n$ sheeted cover
with branch pattern as above, then $g(\widetilde{C} )=n(h-1)+g$ where
$h=g(C)$, dim $P =(n-1)h-n+g$ and $\widetilde{C}$ osculates $P$ to order
$g-1$. We have an abelian soliton if $g-1\geq (n-1)h-n+g$ i.e. 
$(n-1)(h-1)\leq 0$, so either $n=1$ or $h=1$ again.

\bigskip
\noindent {\bf \S 2 Integrable systems}
\medskip
{\bf 2.1 Theorem.}  {\it For every $n \geq g \geq 2$, the family of
degree $n$, genus $g$ coelliptic solitons constructed in {\bf
\S 1.4} is birationally equivalent to
%admits generically the structure of 
an algebraically completely integrable system with
$g-1$ commuting Hamiltonians and $(g-1)$-dimensional Prym varieties $P$ as
Liouville tori.}
\medskip
%{\bf Proof.}
We prove the theorem by identifying our family of coelliptic solitons
as the symplectic reduction of a well-known aci
%algebraically integrable 
system, namely Markman's system of meromorphic Higgs bundles. We start
in ${\bf \S 2.1}$ by reviewing Markman's system. Its total space $X$
parametrizes semistable meromorphic Higgs bundles $(E,\phi)$ on a curve
$C$ with given structure group $G$ (which we take to be either $GL(n)$
or $SL(n)$), polar divisor $D \subset C$, and conjugacy class $O$ of the
residue $Res_{D}(\phi)$ of $\phi$ along $D$. The base $B$ parametrizes
appropriate spectral covers $\widetilde{C}\rightarrow C$, and the
Hamiltonian map $H: X \to B$ sends a meromorphic Higgs bundle $(E,\phi)$
to its spectrum $\widetilde{C}$ defined by the characteristic polynomial
of $\phi$.

For a general Hamiltonian function $h: B \to {\bf C}$ (or a collection
of commuting Hamiltonians $h: B \to {\bf C}^k$), it is {\it not} possible to
reduce a given algebraically integrable system $H: X \to B$ to one whose
base is the hypersurface (or subvariety) $\{ h=0 \} \subset B$. We
explain this difficulty in ${\bf \S 2.2}$, and analyze the
conditions under which such symplectic reduction {\it can} be done in our      
algebraic setting.

In ${\bf \S 2.3}$ we construct (generically)
an embedding of our family  of
coelliptic solitons into Markman's system. The image satisfies the
conditions of ${\bf 2.2}$, so it can be identified with a symplectic
reduction of Markman's system, completing the proof.

\bigskip
\noindent{\bf \S2.1 Markman's system}        
\bigskip

Let $C$ be a (complete, non-singular) algebraic curve with canonical
bundle $\omega_{C}$, and let $D$ be an effective divisor on $C$. A
meromorphic Higgs bundle on $C$ is a pair $(E,\phi)$ where $E$ is a rank
$n$ vector bundle on $C$ and $\phi: E \to E \otimes \omega_{C}(D)$ is an
endomorphism of $E$ taking values in meromorphic differentials on $C$ with
poles on $D$. There is a moduli space ${\cal H}iggs^n_{C,D}$ parametrizing
Higgs bundles on $C$ satisfying an appropriate stability condition, cf.
[M, DM1].

A Higgs bundle $(E,\phi)$ determines a spectral cover $\widetilde{C}$, 
namely the
subscheme of{\hfil\break} 
${{Tot}}(\omega_{C}(D))$ defined by the characteristic
polynomial of $\phi$, where ${{Tot}}(\omega_{C}(D))$ is the total
space of the line bundle $\omega_{C}(D)$ on $C$. The natural projection
of ${{Tot}}(\omega_{C}(D))$ restricts to a map of degree $n$,
$\pi:\widetilde{C} \to C$. The Higgs bundle determines also a sheaf $L$
on $\widetilde{C}$, generically a line bundle, parametrizing the
eigenspaces of the endomorphism $\phi$. Conversely, the Higgs bundle
$(E,\phi)$ can be recovered from the spectral data $(\widetilde{C},L)$:
$E$ is the direct image $\pi_{*}(L)$, while $\phi$ is $\pi_{*}$ of
multiplication by the tautological section, cf. [BNR,M]. In particular,
there is a well-defined morphism $H: {\cal H}iggs^n_{C,D} \to B$, where
$B$ is the vector space parametrizing spectral covers, and the fiber
over a generic $\widetilde{C}$ is {{Pic}}$(\widetilde{C})$.

The crucial point, proved in [M], is that this is a Poisson integrable
system. In particular,  ${\cal H}iggs^n_{C,D}$ has a Poisson structure,
so it is foliated by (algebraically) symplectic submanifolds. The
restriction of $H$ to each such symplectic submanifold makes it into an
integrable system in the usual, symplectic, sense. 

In fact, Markman gives an explicit description of these symplectic
leaves. In case the polar divisor $D$ is reduced, such a leaf is
determined by specifying the conjugacy classes in the Lie algebra
$gl(n)$ of the residues ${{Res}}{\phi}$ of $\phi$ at each point $q
\in D$. (When $D$ is not reduced, one must consider instead coadjoint
orbits in more complicated Lie algebras.)

There is a variant of the meromorphic Higgs bundles for each reductive
group $G$: the vector bundle $E$ is replaced by a $G$-bundle, and $\phi$
is now in $H^0(C,ad(E) \otimes \omega_C(D))$. The case $G=GL(n)$ is the
one above. For $G=SL(n)$, we get a subsystem of the above given by two 
conditions: the trace of $\phi$ (i.e. the first coefficient in the 
characterstic
polynomial of $\phi$) must vanish, and the norm of $L$ must be the
trivial line bundle ${\cal O}_C$, i.e. $L$ must be in
Prym$({\widetilde{C}}/C)$ instead of Pic$(\widetilde{C})$.

In case $C$ is elliptic, we can identify $\omega_C$ with ${\cal O}_C$.
If $D$ consists of $d > 0$ distinct points, it is easy to see
explicitly that the number of Hamiltonians for the $GL(n)$ system (i.e.
the dimension of the base $B$) is $dn(n+1)/2$. Of these, $dn-1$ are
Casimirs, while the remaining $1+dn(n-1)/2$ Hamiltonians induce non-zero
flows on generic symplectic leaves. The generic spectral cover
$\widetilde{C}$ is of genus $1+dn(n-1)/2$, and the generic symplectic
leaf has dimension $2+dn(n-1)$ and is an algebraically integrable
system. The total space has dimension $1+dn^2$.

Indeed, the equation of a spectral cover $\widetilde{C}$ is of the form:
$$\sum_{i=0}^n a_i t^{n-i}=0,$$
where $a_i$ is an arbitrary element of the $di$-dimensional vector space 
$H^0(C, {\cal O}_C(iD))$ for $i>0$, and $a_0=1$. This gives the number of 
Hamiltonians. The $n$-sheeted
covering map $\widetilde{C} \to C$ is ramified at $dn(n-1)$ points,
giving the genus. In terms of the Higgs field $\phi$, the Casimirs are the
symmetric functions of the residues $Res_q(\phi)$ at points $q \in D$. 
There are thus $dn$
of them, subject to one relation coming from the residue theorem; hence
the $dn-1$ independent Casimirs. These Casimirs can also be seen in terms
of the coefficients $a_i$: they are just the leading terms, or images of
the $n$  $a_i$'s in $H^0(C, {\cal O}_C(iD))/H^0(C, {\cal O}_C((i-1)D))$.
For $i \geq 2$ this is the same as the $d$-dimensional space
$H^0(D, {\cal O}_C(iD))$, but for $i=1$ it is a codimension $1$ subspace of 
$H^0(D, {\cal O}_C(D))$.

For the $SL(n)$ system, these dimensions are slightly modified: the
dimension of $B$ is now $d(n-1)(n+2)/2$ (the drop of $d$ comes from
setting $a_1=0$); the number of Casimirs drops by $d-1$, to $d(n-1)$; and the
dimension of the Liouville tori drops by only $1$: Pic$(\widetilde{C})$
is replaced by Prym$({\widetilde{C}}/C)$. The dimension of a generic
symplectic leaf drops by $2$ to become $dn(n-1)$, and the dimension of
the total space drops by $d+1$, to $d(n^2-1)$.

\vfil\eject
\noindent{\bf \S2.2 Symplectic reduction in an algebraic setting}
\bigskip

In general, it is not possible to reduce a given algebraically
completely
integrable system to a smaller one. This is precisely for the same 
reason that not every KP soliton is an abelian soliton: a general
straight line flow on an abelian variety is ergodic, so it is not
tangent to any abelian subvariety.

Let $H:X \to B$ be an aci
%algebraically integrable 
system, and $h:B \to {\bf
C}$ a Hamiltonian function, determining a hypersurface $B_0:=\{ h=0 \}
\subset B$ and a straight line flow along the Liouville torus $A_b$ 
above each point $b \in B_0$. Locally we {\it can} always find a symplectic
reduction, e.g. by intersecting $H^{-1}(B_0)$ with another hypersurface
transversal to the flow. But even if we can form this symplectic reduction 
globally, it will not usually give a reduced 
algebraically
integrable system with 
base $B_0$. The problem is that this second hypersurface 
will intersect each Liouville torus $A_b$ in a hypersurface which is
unlikely to be a subtorus; a generic abelian variety does not contain
{\it any} proper subtori.

So, we want to know when there is an aci
%integrable 
system whose base is 
$B_0$; the general fiber $\overline{A}_b$ should be an abelian variety,
the quotient of $A_b$ by the flow. This is possible in two cases:

\bigskip\noindent (1) The general $A_b$ is an abelian variety, isogenous 
to a product $\overline{A}_b \times C$ for some abelian variety 
$\overline{A}_b$ and elliptic curve $C$; the flow is tangent to $C$.

\bigskip\noindent (2) The general $A_b$ is a ${\bf C}$ or 
${\bf C}^*$-extension of an abelian variety $\overline{A}_b$, and the flow 
is along the ${\bf C}$ or ${\bf C}^*$ fiber.

\bigskip

Both situations do arise. As an example of (1), consider Markman's system 
for $GL(n)$. The Liouville tori are the Jacobians of the spectral covers, 
Jac$(\widetilde{C})$, and are isogenous to Prym$({\widetilde{C}}/C)
\times C$. Let $h$ be any linear function of $a_1 \in H^0(C,{\cal
O}_C(D))$, viewed as a function on $B$. It is a Casimir if and only if it 
is zero on the $1$-dimensional subspace $H^0(C,{\cal O}_C) \subset 
H^0(C,{\cal O}_C(D))$. If it is not, it induces a non-zero flow which is
tangent to the elliptic factor $C$. The symplectic reduction is an
algebraically integrable system, in fact it is just Markman's system for
$SL(n)$. 

Let us consider the case that the system $H:X \to B$ consists of the Jacobians 
of a family of curves Jac$(\widetilde{C_b}), b \in B$, 
as in the $GL(n)$ Markman 
system. When the curve $\widetilde{C_b}$ becomes singular, the fiber of the 
integrable system can become either the generalized Jacobian 
Jac$(\widetilde{C_b})$,
which parametrizes all line bundles on $\widetilde{C_b}$, or perhaps some 
(partial) 
compactification such as the compactified Jacobian of [AK] which parametrizes 
all 
rank one torsion free sheaves on $\widetilde{C_b}$. We will use only the open 
and non-singular part, Jac$(\widetilde{C_b})$.
A 
{\it nonlinear}
reduction of type (2) arises if we take $h:B \to {\bf C}$ to be the
discriminant of $\sum_{i=0}^n a_i t^{n-i}=0,$ the polynomial on $B$ which
vanishes on the hypersurface $B_0 \subset B$ parametrizing singular
spectral covers. For general $b \in B_0$ the spectral cover
${\widetilde{C_b}}$ has a node or a cusp, so its normalization is a curve 
${\widetilde{C_b}}'$ of genus $g'=g-1$, where $g=1+dn(n-1)/2$ is the
genus of the generic spectral cover (over a point not in $B_0$). It
follows that the (generalized, uncompactified) Jacobian, 
Jac$({\widetilde{C_b}})$,
is a  bundle over the ordinary Jacobian Jac$({\widetilde{C_b}}')$ of the 
normalization, with fiber 
${\bf C}^*$ or ${\bf C}$. The Hamiltonian vector field corresponding to this 
$h$ 
is tangent to the ${\bf C}^*$ or ${\bf C}$ fibers. So the symplectic 
reduction exists globally; it is an
algebraically symplectic space, fibered over the discriminant hypersurface 
$B_0$ with general fiber Jac$({\widetilde{C_b}}')$. The general result
is:

\bigskip{\bf 2.2 Theorem.} {\it Let $X \to B$ be an
algebraically 
completely integrable system for which $X$ is algebraically symplectic, 
the dimension of $B$ is $\widetilde{g}$, and the generic fiber $A_b$ is
the Jacobian of a non-singular curve ${\widetilde{C_b}}$ of genus
$\widetilde{g}$. Let $B_g$ be a $g$-dimensional irreducible component of the
subvariety of $B$ parametrizing curves of geometric genus $\leq g$, and
let $B_g^0 \subset B_g$ be an open subset which is non-singular and
such that for $b \in B_g^0$ the curve ${\widetilde{C_b}}$ is integral
(=reduced and irreducible) of geometric genus $g$. Then the family of
Jacobians ${\rm Jac}(\nu({\widetilde{C_b}}))$ of the normalizations 
$\nu({\widetilde{C_b}})$ of the curves ${\widetilde{C_b}}$ for 
$b \in B_g^0$  inherits from $X$ the structure of an algebraically
completely integrable system.}
\medskip
%\bigskip\noindent
{\bf Proof.}
Inductively, we find a nested sequence of
subvarieties:

$$B_g \subset B_{g+1} \subset \ldots \subset B_{\widetilde{g}} = B$$

\noindent such that each $B_i$ is irreducible, $i$-dimensional, is a
component of the subvariety of $B$ parametrizing curves of geometric
genus $\leq i$, and contains a non empty open subset $B_i^0$ which is 
non-singular and such that for $b \in B_i^0$ the curve ${\widetilde{C_b}}$ 
is integral of geometric genus $i$. For each $i$ the inclusion $B_i
\subset B_{i+1}$ corresponds to a reduction of type (2) above, so
inductively we find that the locus of Jacobians Jac$(\nu({\widetilde{C_b}}))$ 
of the normalizations $\nu({\widetilde{C_b}})$ of the curves 
${\widetilde{C_b}}$ for $b \in B_i^0$, for each $i$, inherits from $X$ the 
structure of an algebraically completely integrable system as claimed. QED

\bigskip
{\bf Note.} In cases such as Markman's $SL(n)$ system, the two types of 
reduction must be combined: first we reduce the $GL(n)$ system, whose fibers
are Jacobians, to a base $B_g$ parametrizing curves of geometric genus $g$;
then we reduce the Jacobians to Pryms over the traceless locus.

\bigskip
\noindent{\bf \S2.3 Embedding coelliptic solitons in Markman's system}
\bigskip

As we saw in \S2.1, the generic symplectic leaves in Markman's system
are quite large - their dimension is on the order of $n^2$. The coelliptic 
integrable system which we are trying to build is much smaller - it has 
total dimension $2g-2$, with $g \leq n$. We are
going to show that the generic Markman symplectic leaf contains a 
$g$-dimensional
family of integral curves whose normalizations are the genus-$g$,
degree-$n$ coelliptic solitons over a given elliptic curve $C$, occurring once 
each. By Theorem 2.2 then, each such symplectic leaf 
produces an algebraically integrable system structure on the coelliptic
solitons.

Let us first consider the case that $n > g$. We will embed the
coelliptic solitons into a symplectic leaf of Markman's system for the
group $G=SL(n)$ and the polar divisor $D$ consisting of the single point
$q \in C$. Let $\pi: {\widetilde{C}} \to C$ be the degree $n$ map of a
coelliptic soliton. The fiber $\pi^{-1}(q)$ can be written as $gp+S$
where the effective divisor $S=p_1 + \ldots +p_{n-g}$ consists, for
generic ${\widetilde{C}}$, of $n-g$ distinct points $p_i$. The
ramification divisor of $\pi$ can similarly be written as $R=(g-1)p+r_1
+ \ldots +r_{g-1}$ with $g-1$ points $r_j$ distinct from $p$. For
notational convenience we also fix a non-zero differential $dz$ on $C$.
This allows us to identify the structure sheaf ${\cal O}_C$ with the canonical 
sheaf $\omega_C$. Via the residue at $q$ it also gives a trivialization
of the fiber of ${\cal O}_C(q)$ at $q$.

In order to fix the symplectic leaf, we need to choose the conjugacy
class $O_q$ of $Res_q(\phi)$. This amounts to choosing a $g \times g$
conjugacy class $O_p$ at $p$ plus a complex number $k_i$ (=$1 \times 1$ 
conjugacy class) at each $p_i$. For $O_p$ we take a single $g \times g$ 
Jordan block with some complex number $k$ on the
diagonal and $1$'s above. For the $k_j$ we take $n-g$ arbitrary
numbers, distinct from $k$ and from each other, whose sum is $-k$. 
The corresponding spectral covers have above $q \in C$ one 
ramificaton point at $p$, where $g$ sheets come together, plus $n-g$ 
separate sheets at the points $p_j$ of the divisor $S$. Whenever such a 
curve is integral of geometric 
genus $g$, its normalization is automatically a coelliptic soliton. We need 
to prove the converse, namely, that  each coelliptic soliton occurs, and 
exactly once, among the normalizations of these spectral covers. We need:
\medskip
%\bigskip\noindent
{\bf 2.3 Lemma.} {\it For a generic coelliptic soliton 
${\widetilde{C}}$, we have
$h^0({\widetilde{C}},{\cal O}_{\widetilde{C}}(gq))=1.$}
\medskip
{\bf Proof.}
Since $\omega_{\widetilde{C}} = {\cal O}_{\widetilde{C}}(R) =
{\cal O}_{\widetilde{C}}((g-1)p+r_1+ \ldots +r_{g-1})$,
the claim follows via Riemann-Roch:

$$h^0({\widetilde{C}},{\cal O}_{\widetilde{C}}(gq))=
1+h^0({\widetilde{C}},{\cal O}_{\widetilde{C}}(r_1+ \ldots +r_{g-1}-p))$$

\noindent from the fact that on a generic ${\widetilde{C}}$ the line
bundle ${\cal O}_{\widetilde{C}}(r_1+ \ldots +r_{g-1})$ has a unique 
global section, which vanishes at the $r_i$ but not at $p$.
QED
\bigskip

Now a map of ${\widetilde{C}}$ to $Tot({\cal O}_C(q))$ compatible with the 
cover map $\pi$ is given by a section 
$f \in H^0({\widetilde{C}}, \pi^{*}({\cal O}_C(q)))$.
We claim that there is a unique section $f$ such that:

\noindent {$\bullet$} The image curve is traceless, i.e. it is a spectral
curve in the $SL(n)$ subsystem.

\noindent {$\bullet$} In terms of our trivialization of the fiber of 
${\cal O}_C(q)$ at $q$, this map sends the points $p_j$ to $k_j$. (It 
then automatically sends $p$ to $k$, by the residue theorem. In
particular, it separates the points in the fiber of $\pi$ over $q$. So
if the original ${\widetilde{C}}$ is integral, so will its image be in 
$Tot({\cal O}_C(q))$ .)

Consider the short exact sequence

$$ 0 \to {\cal O}_{\widetilde{C}}(gp) \to \pi^{*}({\cal O}_C(q)) \to 
{\cal O}_S \to 0, $$

\noindent where again we used the differential $dz$ to trivialize the
last sheaf. The long exact sequence gives:

$$0 \to H^0({\widetilde{C}}, {\cal O}_{\widetilde{C}}(gp))
\to H^0({\widetilde{C}}, \pi^{*}({\cal O}_C(q)))
\to H^0(S, {\cal O}_S) 
\to H^1({\widetilde{C}}, {\cal O}_{\widetilde{C}}(gp)).$$

\noindent By the Lemma, the first term is $1$-dimensional. By Riemann-Roch 
then, the
last term vanishes. So for every specification of residues $\{k_j,
j=1,\ldots ,n-g \} \in H^0(S, {\cal O}_S)$ there is a $1$-
dimensional affine space of liftings 
$f \in H^0({\widetilde{C}}, \pi^{*}({\cal O}_C(q)))$
with these specified residues. It remains to show that among them there
is a unique one which is traceless. Equivalently, we need that the trace
map:

$$Tr: H^0({\widetilde{C}}, \pi^{*}({\cal O}_C(q))) \to H^0(C,{\cal O}_C(q))$$

\noindent is non-zero on the one dimensional subspace
$H^0({\widetilde{C}}, {\cal O}_{\widetilde{C}}(gp))
\subset H^0({\widetilde{C}}, \pi^{*}({\cal O}_C(q)))$.
But the trace map can be written more naturally as:
$$Tr: H^0({\widetilde{C}},\omega_{\widetilde{C}}(S+p-\sum r_i))
\to H^0(C,\omega_C(q)).$$
Our one dimensional subspace 
$H^0({\widetilde{C}}, {\cal O}_{\widetilde{C}}(gp))$
is then identified with $\pi^{*}$ of $H^0(C,\omega_C)=H^0(C,\omega_C(q))$. 
On this the trace map is simply multiplication by $deg(\pi)=n$, so it is 
non-zero as required.

Now we need to consider the remaining case, $n=g$. In this case the
previous divisor $S$ is empty, so $H^0({\widetilde{C}}, {\cal 
O}_{\widetilde{C}}(gp))
\to H^0({\widetilde{C}}, \pi^{*}({\cal O}_C(q)))$ is an isomorphism,
hence the only traceless section $f$ is $f=0$. But the resulting curve
is not integral: it is the $0$-section $C$ with multiplicity $g$.

Instead, we consider an $SL(n)$ Markman system with a higher polar
divisor. For example, $D=q+q'$ will do, where $q'$ is any point of $C$
other than $q$. We have $\pi^{-1}(q)=gp$, and we set $S:=\pi^{-1}(q')$.
In this new notation we still have an exact sequence

$$0 \to H^0({\widetilde{C}}, {\cal O}_{\widetilde{C}}(gp))
\to H^0({\widetilde{C}}, \pi^{*}({\cal O}_C(q+q')))
\to H^0(S, {\cal O}_S) 
\to H^1({\widetilde{C}}, {\cal O}_{\widetilde{C}}(gp)).$$

\noindent By the Lemma, the first term is $1$-dimensional and the last term
vanishes. The trace map

$$Tr: H^0({\widetilde{C}}, \pi^{*}({\cal O}_C(q+q'))) \to H^0(C,{\cal 
O}_C(q+q'))$$

\noindent can now be identified with:

$$Tr: H^0({\widetilde{C}},\omega_{\widetilde{C}}(S+p-\sum r_i))
\to H^0(C,\omega_C(q+q')).$$

\noindent The $1$-dimensional subspace 
$H^0({\widetilde{C}}, {\cal O}_{\widetilde{C}}(gp))$
is identified with 
$$\pi^{*}H^0(C,\omega_C) \subset \pi^{*}H^0(C,\omega_C(q+q')).$$
So as before, the trace map on it is multiplication by $n=g$, hence 
non-zero. So a generic choice of residue in $H^0(S, {\cal O}_S)$ will indeed
lift to a unique traceless section $f$, producing the desired embedding.
QED

{\bf Remarks. (1)}
We defined an aci whose integral manifolds consist 
of Jacobians (or Prym varieties), so the results of \S7.2 in [DM1]
predict the existence of a cubic $c\in H^0(B, Sym^3{\cal V})$,
where $B$ is the `base' (the parameter
space of $g-1$ points on a fixed elliptic curve)
and ${\cal V}$ is the vertical tangent bundle of
the Prym fibration.
This too will be the reduced version of the Hitchin-Markman cubic
which is identified in [DM1] and [DM2].
%The cubic condition is analytic, so even though we had
%to make a genericity assumption at several turns during our construction,
%the aci property will have to hold on the closure of the
%manifold we considered, as long as we take the
%ambient space to be an algebraic fibration, such as the
%set of all co-elliptic solitons over a given elliptic
%curve with a fixed point $p$, and the attendant Pryms as fibres.

{\bf (2)} As in [DM1 \S6], we could ask whether the
KP flows on the fibres are compatible with the 
Poisson structure of our constrained Hitchin-Markman system.
The answer is the same as in [DM1] (in the affirmative, that is):
since the Poisson structures were compatible before reduction, they
can be both reduced and remain compatible.

\bigskip
\noindent{\bf Appendix}
\medskip
The simplest examples of abelian solitons would be
fibered products $\widetilde{C} =C \times_{{\bf P}^1}C^\prime$ where
$\widetilde{C}\rightarrow C^\prime$ has a branch point of order $h$ at $p$ and
Jac $\widetilde{C}\sim$ Jac $C\oplus$ Jac $C^\prime$. Indeed, we show (A.2)
by using results of Treibich [T] that
these are the only 
possibilities for tangential covers when genus $C>1$,
however we found it surprising that no such
example may give an abelian soliton (A.3, A.4).
\medskip
 {\bf A.1 Theorem} ([T], 2.5). {\it If $(\widetilde{C}, p)\rightarrow (C,q)$
is a degree $n$ tangential cover with genus $C>1$, then $n=2, C$ and
$\widetilde{C}$ are hyperelliptic, $q$ is a Weierstrass point of $C$ and $p$
is not a Weierstrass point of $\widetilde{C}$.}
\bigskip
{\bf A.2 Corollary.} {\it $(\widetilde{C}, p)\rightarrow (C,q)$ is a tangential
cover with genus $C=h>1\Leftrightarrow\widetilde{C}$ is a fibered product $C
\times_{{\bf P}^1}C^\prime$ with $C^\prime$ of genus zero, branched at say
$\{ 0,\infty\}$, $C$ hyperelliptic branched at $\{
e_1,\ldots,e_{2h+2}\}$ and $C,C^\prime$ have one or two branchpoints
in common.}
\medskip
{\bf Proof.} ``$\Rightarrow$'' Treibich shows that $\widetilde{C}$ is defined
by a spectral polynomial $T^2+bT+\wp$, where $b$ is a constant, $\wp$
is a function on $C$ with a double pole at $q$ and regular elsewhere;
if $k$ is the function in $K(\widetilde{C} )$ whose minimal polynomial is
$T^2+bT+\wp$, then $\widetilde{C}\rightarrow C$ is branched where $4\wp =b^2$
and over $\infty$; however, since both $k$ and $\wp$ are branched over
$\infty$ the fiber product acquires a singularity which we resolve. If
$b^2/4$ is not a branchpoint of $C$, then $k: \widetilde{C}\rightarrow
C^\prime$ is branched at $2(2h+1)$ points and $g(\widetilde{C} )=2h$; if
$b^2/4$ is a branchpoint of $C$, then $g(\widetilde{C} )=2h-1$. Note that
Galois $(\widetilde{C} /{\bf P}^1)$ is the noncyclic group of order 4 and the
third curve $C^{\prime\prime}$ is given by the function 
$$\sqrt{\prod_{i=1}^{2h+1}(\wp -e_i)(\wp -\alpha )},$$
so that Jac $\widetilde{C}\sim$ Jac $C\oplus$ Jac $C^{\prime\prime}$: 
$C^{\prime\prime}$ is not
branched over $\infty$, so $\widetilde{C}\rightarrow C^{\prime\prime}$ is.

\noindent ``$\Leftarrow$'' The general fibered product $\widetilde{C}
=C\times_{{\bf P}^1}C^\prime$, where $\pi :C\rightarrow {\bf P}^1$ is a
double cover branched at ${ S}$: $\{ p_1,\ldots,p_{2 h+2}\}$ and
$\pi^\prime : C^\prime\rightarrow {\bf P}^1$ is a double cover
branched at $S^\prime =\{ 0,\infty\}$, has a third quotient $C^{\prime\prime}$
with $\pi^{\prime\prime}: C^{\prime\prime}\rightarrow 
{\bf P}^1$ branched at the symmetric
difference ${S}\Delta { S}^\prime$, so that
$h^{\prime\prime}=g(C^{\prime\prime})=h+1-\# ({ S}\cap S^\prime )$, 
$g=g(\widetilde{C}
)=h+h^\prime$. The natural decomposition of holomorphic differentials
$H^0(\Omega_{\widetilde{C}} )=H^0(\Omega_C)\oplus H^0(\Omega_{C^\prime})$ gives
a corresponding decomposition of canonical spaces, ${\bf P}^{g-1}:{\bf
P}(H\oplus H^\prime )$. Let $\phi_{\widetilde{C}}$ be the canonical map which
sends $\widetilde{C}$ to a rational normal curve in ${\bf P}^{g-1}$;
projecting from ${\bf P}H$ to ${\bf P}H^\prime$ gives a map
$\widetilde{C}\rightarrow {\bf P}^{h^\prime -1}$ which factors through the
canonical map $\phi_{C^\prime}:C^\prime \rightarrow {\bf P}^{h^\prime
-1}$, and similarly for $\phi_C$. Thus, $\rho :\widetilde{C}\rightarrow C$ is
a tangential cover at $p\in\widetilde{C}\Leftrightarrow\phi_{\widetilde{C}} 
(p)=\phi_C
(\rho (p))\in {\bf P}^{g-1}\Leftrightarrow\phi_{\widetilde{C}} (p)\in {\bf
P}^{g-1}\Leftrightarrow p$ is a base point of the linear system
$\rho^{\prime\ast} H^0(\Omega_{C^\prime})\subset H^0(\Omega_{\widetilde{C}}
)\Leftrightarrow\rho^\prime$ is ramified at $p$ $\Leftrightarrow\pi
(\rho (p))\in S\cap S^\prime$. Cases: \# $(S\cap S^\prime )=1,
h^\prime =h$, there are two such points $p$; \# $(S\cap S^\prime )=2,
h^\prime =h-1$, there are 4 such points $p$; \# $(S\cap S^\prime )=0$,
$\widetilde{C}\rightarrow C$ is not tangential but $\widetilde{C}\rightarrow 
C^{''}$
is. QED

\medskip

{\bf Remark}. $\widetilde{C}$ cannot be the fibered product of more than 2
curves $C,C^\prime, C^{''},\ldots$ with Jac $C$ an abelian soliton, for
this would imply that $\widetilde{C}\rightarrow C^\prime$ and
$\widetilde{C}\rightarrow C^{''}$ are both branched at the same point
$p\in\widetilde{C}$. 
\medskip
 {\bf A.3 Proposition}. {\it Let $\pi_1 :\widetilde{C}_1\rightarrow C$ be an
abelian cover with group $G$ of order $d$ and let $\{ G_i\}_{i\in I}$
be the lattice of subgroups of $G,\ \widetilde{C}_i=J_1/G_i,\ J_i=$ Jac
$\widetilde{C}_i$ and $P_i=J_i/(\Sigma J_k,\ \hbox{\rm for}\ G_i<\not=
G_k)$. Then $J_1\sim\oplus_{i\in I} P_i$, and
$P_1=0\Leftrightarrow J_1=\sum_{i>1}J_1\Leftrightarrow$ either $G$ is
noncyclic or $g(C)=g(\widetilde{C}_1)=g_1$. The latter situation occurs 
$\Leftrightarrow$
$\widetilde{C}_1$ and $C$ are both elliptic and $\pi$ unramified, or both
rational and $\pi$ given by $z\mapsto z^n$, some $n\in {\bf Z}$.}
\medskip
{\bf Proof.} Since $G$ is abelian, $H^0(\Omega_{\widetilde{C}_1})$ decomposes
a $G$-module into 1-dimensional subspaces; the action of $G$ on each
of these factors through a cyclic quotient, so that $J_1\sim\oplus
P_i$ (sum over some $i$'s for which $G/G_i$ is cyclic); in particular,
$P_1\not= 0\Rightarrow G$ cyclic. Now assume $G$ cyclic, so that $I=\{
e\in {\bf Z}_{+}|\ e|d\}$, $|G_e|=e$, deg $(\pi_e :\widetilde{C}_e\rightarrow
C)=|G/G_e|=d/e$. If none of the coverings has any branchpoints, or
equivalently $\pi_1$ is unramified, then $g_e=1+{d\over e}(h-1)$; but
dim $P_e=g_e-\sum_{\scriptstyle{{{f|e}\atop {{f<e}}}}}\dim P_f$ where
the sum is taken over the maximal elements of the lattice; there
may be repetitions arising from chains (when $e$ isn't square-free)
so by induction dim $P_e=(h-1){d\over e}\prod_{p|{d\over e}}(1-{1\over
p})$ for $p$ a prime number. In particular, dim
$P_1=(h-1)d\prod_{p|d}(1-{1\over p})$ is 0 iff $h=1$. If there is one
branchpoint for $\pi_1$, with stabilizer $G_i$ say, then for $e|i$ the
stabilizer of the corresponding point in $\widetilde{C}_e$ is $G_{i/e}$, so
the branching contribution to $g_e$ is ${1\over 2}({i\over
e}-1){d\over i}={1\over 2}({d\over e}-{d\over i})$ and by the same
analysis as above the contribution to $P_1$ is ${d\over
2}\prod_{p|i}(1-{1\over p})>0$. All in all, dim
$P_1=(h-1)d\prod_{p|d}(1-{1\over p})+\sum_{\rm branchpoints} {d\over
2}\prod_{p|i}(1-{1\over p})$. This could be zero again only if $g=0$
and $\pi$ has 2 branchpoints, each totally ramified. QED

\medskip
We end this section by excluding the possibility of an abelian soliton
of the ``simplest'' type, namely: $\widetilde{C}\rightarrow C$ a Galois
cover, Jac $\widetilde{C}\sim$ Jac $C\oplus A$, dim $A=n,\ A=\oplus P_e$
summed over the Pryms of coverings $\rho_e :\widetilde{C}\rightarrow C_e$
which have a total ramification of order $n$ at a given point
$p\in\widetilde{C}$. We believe this to be impossible in general (for $n>1$), 
but
will only address some specific situations for ease of calculation.
\medskip
{\bf A.4 Proposition.} {\it Let $\widetilde{C}\rightarrow C$ be an abelian
cover with Galois group $G$ which is an $\ell$-group of exponent $\ell^n$,
noncyclic
(for some prime number $\ell$). 
Then $\widetilde{C}$ cannot be an abelian soliton with respect to
a branch point $p\in\widetilde{C}$ of index $\ell$.}
%, in the sense that the first
%$k$ osculating planes to $\widetilde{C}$ at $p$ lie on an abelian variety of
%dimension $k$.}
\medskip
{\bf Proof}. Let $h\in G$ be an element of order $\ell$ such that $<h>$
is the stabilizer of $p$. For all $G_e\leq G$, we define
$C_e=\widetilde{C} /G_e,\ \pi_e :C_e\rightarrow C,\ \rho_e 
:\widetilde{C}\rightarrow
C_e$ and $P_e\leq$ Jac $C_e$ to be the Prym of $\pi_e$. If $A=\oplus$
($P_e$ such that $\rho_e$ is not ramified at $p$), then the first
$\ell -1$ osculating planes to $\widetilde{C}$ at $p$ are contained in $A$, so
our condition for abelian soliton is: $\ell -1\geq\dim A$. By the
derivation obtained in the proof of A.3, dim $A={{\ell -1}\over
\ell}\sum_{\scriptstyle{{{h\notin G_e}\atop {{G/G_e\ cyclic}}}}} [G\colon
G_e](g-1+{1\over 2}r_e)$, where $r_e=$ \# branchpoints of $\rho_e$.
Indeed, $P_e=0$ unless $\pi_e$ is cyclic, i.e. $G/G_e$ is cyclic, and
$\rho_e$ is ramified at $p\Leftrightarrow h\not\in G_e$. In general, a
branch point of order $i$ contributes ${1\over
2}[G:G_e]\prod_{{\scriptstyle{{q\ {\rm prime}}\atop  {{q|i}}}}}(1-{1\over
q})$; however, since $G$ is an $\ell$-group all branchpoints contribute
equally. The condition then becomes $2\geq \Sigma_e {{[G:G_e]}\over
\ell}(2g-2+r_e)$ where only integers appear and $r_e\geq 1$. Thus $g\leq
1$, for $g=1$ there can only be one term with $r_e=2$, $[G:
G_e]=\ell$; or one with $r_e=1, [G:G_e]/\ell =2$; or two with $r_e=1,
[G:G_e]/\ell=1$; similarly there are only three possible $r_e$ when $g=0$
and one rules them out case by case. QED
\medskip

The curves constructed in 1.4 generically are not fibered products,
since they have one total ramification point and the remaining
ramifications are simple. When the $g-1$ simple ramifications coincide
and the difference $p_0-p_1$ on the elliptic curve is a $k$ torsion
point ($p_0,p_1$ are the points on $E$ where the map is ramified)
then there is a map $E\rightarrow {\bf P}^1$ of degree $k$, totally
ramified at $p_0$ and $p_1$. If $k,g$ are coprime then $\widetilde{C}$ is the
corresponding fibered product of $E$ and ${\bf P}^1$, but the Galois
group of $\widetilde{C}\rightarrow {\bf P}^1$ is cyclic, so that Jac 
$\widetilde{C}$
does not decompose (A.3).

\bigskip

\noindent{\bf Bibliography}
%\medskip
%\item{[AHH]}
%M.R. Adams, J. Harnad and J. Hurtubise,  Coadjoint orbits, 
%spectral curves and Darboux coordinates, in {\it The
%geometry of Hamiltonian systems} (Berkeley, CA, 1989), 9-21, 
%Math. Sci. Res. Inst. Publ., {\bf 22}, Springer, New York, 1991.

\medskip
\item{[AMcKM]} H. Airault, H.P. McKean and J. Moser, Rational and
elliptic solutions of the KdV equation, {\it Comm. Pure Appl. Math.}
{\bf 30} (1977), 95-148.
\medskip
\medskip
\item{[AK]}
A.B. Altman  and S.L. Kleiman, Compactifying the Jacobian,
 {\it Bull. Amer. Math. Soc.} {\bf 82} (1976), no. 6,
947-949.

%\medskip
%\item{[AG]}
%V.I. Arnol'd and A.B. Givental',  {\it Symplectic geometry},
%Current problems in mathematics. Fundamental directions, 
%Vol. 4, 5--139, 291, Itogi Nauki i
%Tekhniki, Akad. Nauk SSSR, Vsesoyuz. 
%Inst. Nauchn. i Tekhn. Inform., Moscow, 1985. 
%\medskip
%\item{[B]} A. 
% Beauville, Jacobiennes des courbes spectrales et 
%syst\`emes hamiltoniens {\hfil\break}compl\`etement int\'egrables,
%{\it  Acta Math.} {\bf 164} (1990), no. 3-4, 211-235.
\medskip
\item{[BNR]}
A. Beauville, M.S. Narasimhan and S. Ramanan, 
 Spectral curves and the generalised
theta divisor, {\it J. Reine Angew. Math.} {\bf 398} (1989), 169--179.

\medskip
\item{[CPP]} E. Colombo, G.P. Pirola and E. Previato, Density of
elliptic solitons,  {\it J. Reine Angew. Math.} {\bf 451} (1994),
161--169. 
\medskip
\item{[DM1]} 
R. Donagi and E. Markman,  Spectral covers, 
algebraically completely integrable, Hamiltonian systems, and
moduli of bundles, {\it Integrable systems and quantum groups 
(Montecatini Terme, 1993)}, 1-119, Lecture Notes in Math., {\bf 1620},
Springer, Berlin, 1996.

\medskip
\item{[DM2]}
R. Donagi and E. Markman, Cubics, 
integrable systems, and Calabi-Yau threefolds, {\it Proceedings of the
Hirzebruch 65 Conference on Algebraic Geometry (Ramat Gan, 1993)}, 
199-221, Israel Math. Conf. Proc., 9, Bar-Ilan Univ.,
Ramat Gan, 1996.
\medskip
\item{[H]}
R. Hartshorne, {\it Algebraic geometry}, Graduate Texts in Mathematics, No.
52. Springer-Verlag, New York-Heidelberg, 1977.
\medskip
%\item{[K]} A. Krazer, {\it Lehrbuch der Thetafunctionen},
%Teubner, Leipzig, 1903.
%\medskip
\item{[K]} I.M. Krichever, Elliptic solutions of the
Kadomtsev-Petviashvili equation and integrable systems of particles,
{\it Functional Anal. Appl.} {\bf 14} (1980), 282-290.
\medskip
\item{[M]} E. Markman, Spectral Curves and Integrable Systems, 
{\it Compositio Math.} {\bf 93} (1994), no. 3, 255-290.
\medskip
\item{[Mu]} D. 
Mumford, {\it Curves and their Jacobians}, The University of Michigan Press, 
Ann Arbor, Mich., 1975.

\medskip
\item{[SW]} G. Segal, and G. Wilson,  Loop groups and equations of 
KdV type, {\it Inst. Hautes \'Etudes Sci. Publ.
Math.} No. {\bf 61} (1985), 5--65.

%\medskip
%\item{[S]} T. Shiota, Characterization of Jacobian varieties 
%in terms of soliton equations, {\it Invent. Math.} {\bf 83} (1986),
%no. 2, 333-382. 
\medskip
\item{[T]} A. Treibich, Tangential polynomials and elliptic solitons,
{\it Duke Math. J.} {\bf 59} (1989), 611-627.
\medskip
\item{[TV1]} A. Treibich and J.-L. Verdier, Solitons elliptiques, {\it
The Grothendieck Festschrift}, Vol. III, Birkh\"auser, Boston, 1990,
437-480.
\medskip
\item{[TV2]} A. Treibich and J.-L. Verdier, Vari\'et\'es de Kritchever
des solitons elliptiques de KP, {\it Proceedings of the Indo-French 
Conference on Geometry} (Bombay, 1989), 187-232,
Hindustan Book Agency, Delhi, 1993.

\baselineskip10pt
\bigskip \noindent {\smallsmc
        Ron Y. Donagi,
        Department of Mathematics,
        University of Pennsylvania,
        Philadelphia, PA 19104-6395
        USA
\par}

donagi@math.upenn.edu

\bigskip \noindent {\smallsmc
        Emma Previato,
        Department of Mathematics and Statistics,
        Boston University,
        Boston, MA 02215-2411
        USA
\par}

ep@math.bu.edu

\end
\bye